\documentclass[12pt,a4paper]{amsart}
\usepackage{amsmath, amssymb, amsfonts, amsthm}
\usepackage{hyperref}
\usepackage[english]{babel}
\usepackage{amsmath}
\usepackage{amssymb}
\usepackage{url}
\usepackage{amsthm}
\usepackage{amsxtra}
\usepackage{mathrsfs}
\usepackage{fancyhdr}
\usepackage{enumerate}
\usepackage{graphicx}
\usepackage{hyperref}

\newtheorem{theorem}{Theorem}

\newtheorem{lemma}{Lemma}[section]

\newtheorem{definition}[lemma]{Definition}

\theoremstyle{definition}

\numberwithin{equation}{section}

\newcommand{\pd}[2]{\frac{\partial {#1}}{\partial {#2}}}
\newcommand{\beq}{\begin{equation}}
\newcommand{\eeq}{\end{equation}}
\newcommand{\be}{\begin{equation*}}
\newcommand{\ee}{\end{equation*}}
\newcommand{\n}{\noindent}

\newcommand{\RE}{\mathbb R}
\newcommand{\erre}{\mathbb R}
\newcommand{\CO}{\mathbb C}

\newcommand{\DD}{\mathscr D}

\newcommand{\HH}{\mathbb{H}}
\newcommand{\GG}{\mathcal{G}}

\newcommand{\lf}{\left}
\newcommand{\ri}{\right}
\newcommand{\ve}{\varepsilon}
\newcommand{\al}{\alpha}

\newcommand{\de}{\delta}

\newcommand{\ome}{\omega}
\newcommand{\Imm}{\mathfrak{Im}}

\newcommand{\Rea}{\mathfrak{Re}}

\newcommand{\tr}{\mathscr{T}}
\newcommand{\rf}{\mathscr{R}}

\DeclareMathOperator{\sech}{sech}

\renewcommand{\leqslant}{\leq}
\renewcommand{\geqslant}{\geq}

\newcommand{\f}{\frac}
\newcommand{\EE}{\mathcal E}

\title[]{Nonlinear Schr\"odinger equation on graphs:\\ recent results and open problems}

\author[]{Diego Noja}
\address{Dipartimento di Matematica e Applicazioni, Universit\`a
 di Milano Bicocca,  via R. Cozzi, 53, 20125 Milano, Italy}
\email{diego.noja@unimib.it} 

\date{}

\begin{document}


\begin{abstract}
In the present paper an introduction to the new subject of nonlinear dispersive hamiltonian equations on graphs is given. The focus is on recently established properties of solutions in the case of nonlinear Schr\"odinger equation. Special consideration is given to existence and behaviour of solitary solutions. Two subjects are discussed in some detail concerning NLS equation on a star graph: the standing waves of NLS equation on a graph with a $\delta$ interaction at the vertex; the scattering of fast solitons through an Y-junction in the cubic case. The emphasis is on description of concepts and results and on physical context, without reporting detailed proofs; some perspectives and more ambitious open problems are discussed.
\end{abstract}

\maketitle

\section{Introduction}
In the last decades a large amount of work has been done concerning existence and behaviour of solutions of nonlinear dispersive equations of Hamiltonian type.
This is in part a consequence of the fact that many fundamental physical models belong to this family. In particular nonlinear Klein-Gordon equations and their relatives are milestones of classical and quantum field theory; but in fact an important boost to these developments comes from more phenomenological models, such as Korteweg-deVries equation describing shallow water waves in certain approximations or the ubiquitous nonlinear Schr\"odinger equation which describes electromagnetic pulse propagation in nonlinear (Kerr) media, Langmuir plasma waves, or in the quantum realm Bose-Einstein condensates, where it is better known under the name of Gross-Pitaevskii equation. All the above Hamiltonian equations share a common characteristic: they admit solitary solutions, or briefly solitons. As it is well known, solitons are solutions emerging due to a balance between nonlinearity and dispersion and they are related to symmetries of the equations; in some relevant examples (as one dimensional KdV and cubic nonlinear Schr\"odinger equation) they also have a strong relation to the complete integrability of the infinite dimensional Hamiltonian system to which they refer, but their existence and main interesting properties are by no means restricted to integrable equations.
Solitons are nondispersive and to some extent particle-like solutions of certain PDEs; similarly to the equilibrium points of finite dimensional dynamical systems, they are an essential point of departure in the description of the phase portrait of the model equation which they solve.
In the present paper it will be shown the existence and role of solitons for some model PDEs on ramified structures, in particular the simplest kind of them, the so called star graphs. The model equation described in some detail is the nonlinear Schr\"odinger equation on graphs, about which a rigorous mathematical activity is developing. Before describing mathematical models let us give a look at the literature of physical origin which motivates the study. Two main fields where NLS equation enters as a preferred model are optics of nonlinear Kerr media and dynamics of Bose Einstein condensates. Both of these quite different physical situations have potential or actual application to graphlike structures. In nonlinear optics we can mention arrays of planar self focusing waveguides, and propagation in variously shaped fiber optics devices, such as Y-junctions,  H-junctions and more complex examples can be considered. Papers relevant to these items are \cite{[BS],[HTM]}, where also symmetry breaking phenomena related to geometry are studied. In \cite{[Linzon],[Pe]} experimental evidence of interaction of solitons with inhomogeneities and defects is given, in particular scattering and capture of solitons in photonic traps and escaping of solitons from potential wells. These last phenomena are studied on one dimensional media but they suggest a natural generalization to simple graphlike structures, such as Y-junctions. In \cite{[GSD]}, after an analysis of general issues discussed later in this paper, an example of potential application to signal amplification in resonant scattering on networks of optical fibers is given. In the field of Bose-Einstein condensates and more generally of nonlinear guidance of matter waves there has been an increasing interest in one dimensional or graphlike structure. Boson liquids or condensates can be treated in the presence of junctions and defects in analogy with the Tomonaga-Luttinger fermionic liquid theory, with applications to boson Andreev-like reflection, beam splitter or ring interferometers (see for example \cite{[TOD],[ZS]} and references therein). Apart experimental activity, some other theoretical and numerical studies should be mentioned. NLS equation on graphs has discrete analogues (see \cite{[BCSTV],[KFTK],[Miro],[SMHM]}), where several dynamical behaviours can be studied analytically and numerically. Other discrete models are spin models on star graphs related to Kondo model (see \cite{[CT]}). Possible integrability of cubic NLS on star graphs with special boundary conditions is discussed in \cite{[Sob]}) and finally quantum field theory on star graphs  with a  consideration of symmetry, integrability and analysis of special models is studied in \cite{[BM]}.

\subsection{Preliminaries and the mathematical model}
To pose the models, we briefly recall some preliminary notions (see\cite{[BCFK06],[EKK08],[BEH],[Kuc04],[Kuc05],[KS99]} for a more systematic account). By a metric graph $\GG$ it is meant a set of edges $\{e_j\}_{j=1}^n$ and vertices, with a metric structure on any edge. Every edge of the graph is identifyied with a (bounded or unbounded) oriented   segment, $e_j \sim I_j$.
A function on a graph is a vector 
\[
\Psi = (\psi_1, . . . ,\psi_N)
\qquad 
\text{with} 
\qquad
\psi_j \equiv \psi_j(x_j)\;; \quad x_j \in I_j\ .
\]
In the following, as above, we will denote the elements of $L^2(\GG)$ by capital Greek letters,
while functions in $L^2(\RE^+)$ are
 denoted by lowercase Greek letters.
The spaces $ L^p(\GG)$ are defined in the natural way by 
$$
 L^p(\GG)= \bigoplus_{j=1}^N L^p(I_j)\;; \quad
 \|\Psi\|_p = \left(\sum_{j=1}^N \|\psi_j\|_{L_p(I_j)}^p\right)^{\frac{1}{p}} \ .
$$
Only for the $L^2$-norm we
drop the subscript and simply write $\| \cdot \|$. Accordingly, we
denote by $(\cdot,\cdot)$ the scalar product in $L^2(\GG)$. 
When an element of $L^2 (\GG)$ evolves in time, we use in notation the
subscript $t$: for instance, $\Psi_t$. Sometimes we shall write $\Psi(t)$ in order
to emphasize the dependence on time, or whenever such a notation is more
understandable.
In a similar way one can define the Sobolev spaces $$H^p(\GG) =  \bigoplus_{j=1}^N H^p(I_j)\ .$$ The definition however should be used with caution, because as such, without specification of behaviour at the vertices, these spaces have not the usual properties of Sobolev spaces, in particular continuous or compact embeddings. The main example of metric graph that will be considered here is a {\it star graph}, which is characterized by a single vertex $v$ and $N$ infinite edges, and we put $e_j \sim (0,+\infty)$, $v\equiv 0 .$
\par\noindent 
After functional spaces, differential operators can be given on graphs. In particular we are interested in operators connected with the Laplacian and some variants.
On the Hilbert Space
\[\HH=\bigoplus_{j=1}^{N} L^2((0, \infty))\]
with elements 
\[\Psi=(\psi_1,\dots,\psi_N)\in\HH\]
endowed by a norm 
\begin{equation*}
\|\Psi\|=
\bigg(\sum_{j=1}^N\|\psi_j\|_{L^2(\RE^+)}^{2}\bigg)^{1/2}
\end{equation*}

\par\noindent
we consider the operator $H_\GG$
\[H_\GG \Psi=\left(-\frac{d^2\psi_1}{dx_1^2},\dots,-\frac{d^2\psi_N}{dx_N^2}\right)\]
on a suitable domain
\[\DD(H_\GG)=\bigoplus_{j=1}^{N} H^2((0,\infty))\ \ \& \ \
\text{self-adjoint conditions at the vertex}
\]
The choice of the boundary condition of course qualifies the operator, and different boundary condition give rise to different dynamics when the operator $H_\GG$ is the generator of an evolution, e.g. Schr\"odinger or Klein-Gordon.



As it is well known, unitary $N\times N$ matrices $U$ parametrize the family of selfadjoint Laplacians on $\GG$, and the relation between $U$ and the specific boundary condition at the vertex is given by

\[(U-1)\begin{pmatrix}
    \psi_{1}(0)\\
    \vdots\\
    \psi_{N}(0)
   \end{pmatrix}+
i(U+1)\begin{pmatrix}
    \psi'_{1}(0)\\
    \vdots\\
    \psi'_{N}(0)
   \end{pmatrix}=0 \]


However we are not interested here in the greatest generality, and we restrict ourselves to the following simple cases:\par\noindent
notice that for a graph with two
edges, i.e. the line, continuity of wavefunction and its derivative
for an element of the domain makes the interaction disappear;
this fact justifies the name of free Hamiltonian.
\par\noindent
-  $\delta$ condition: [$U_{jk}=2(N+i\alpha)^{-1}-\delta_{jk}$]
\[
\psi(v)\equiv\psi_1(0)=\psi_2(0)=\dots=\psi_N(0)
\;,\qquad\sum_{j=1}^N\psi'_j(0)=\alpha\psi(0) \qquad \alpha\in\RE
\]
The $\delta$ condition includes the so called Kirchhoff condition $\alpha=0$ as a special case, but it is convenient to distinguish between them. Notice that for a graph with two edges, i.e. the line, continuity of wavefunction and its derivative
for an element of the domain makes the interaction disappear; this fact justifies the name of free Hamiltonian.
\par\noindent
From now on $H_{\alpha}$ is the operator $H_\GG$ with delta condition in the vertex with ``strength'' $\alpha$. To simplify notation the Kirchhoff case will be indicated with the symbol $H$. Another more singular interaction to which we will refer occasionally is given by the \par\noindent  
- $\delta^{'}_s$ condition:
\[
\sum_{j=1}^n \psi_j^\prime(0)=0 \
,\quad
\psi_j(0) -\psi_k (0) =\frac{\beta}{n}(\psi_j^\prime(0) -\psi_k^\prime(0))\
,\quad j,k= 1,2,...,n\ ,
\]
which coincides, in the case of the line, with a $\delta'$ interaction of strength $\beta/2$.
The operator $H_{\alpha}$ has simple spectral characteristics. The absolutely continuous spectrum coincides with $[0,\infty)\ .$ As regards the point spectrum,\\
- For $\alpha<0$, $\sigma_{p}(H_{\alpha}) = -\alpha^2/N^2$\\
- For $\alpha\geq0$, $\sigma_{p}(H_{\alpha}) =\{\emptyset\}$ \\
The presence/absence of the (negative) eigenvalue is the motivation for the name of attractive/repulsive delta vertex given to $H_{\alpha}$ in correspondence to the cases $\alpha<0$ $/$ $\alpha>0$. More specifically, a $\delta$ vertex
with $\alpha<0$ can be interpreted as modeling an
attractive potential well or attractive impurity. In fact, as in the case of the line, the
operator $H_{\alpha}$ is a norm resolvent limit for $\epsilon$ vanishing of a
scaled Hamiltonian $H_\epsilon=H+\alpha V_\epsilon$, where
$V_\epsilon=\frac{1}{\epsilon}V(\frac{x}{\epsilon})$ and $V$ is a
positive normalized potential defined on the graph in the natural way and $H$  is the free Hamiltonian (\cite{[BEH]} and
reference therein). 
\par\noindent
Finally, the quadratic form associated to $H_{\alpha}$ is
\[
 Q[\Psi] = \frac12 \|\Psi'\|^2 + \frac\alpha2 |\psi(0)|
\]
\[
 \DD(Q) = \{\psi\in H^1\; \text{s.t.}\; \psi(0)\equiv \psi_1(0)=...=\psi_N(0)\}\equiv \EE
\]
and 
\[
 \Psi' \equiv (\psi_1', ... ,\psi_N')^T\ .
\]
\par\noindent
Notice that $\DD(Q)$, the form domain of $Q$, is independent on $\alpha$. In the following we will use the convenient notation $\EE$, calling it the {\it energy } domain (it is often considered the "true" Sobolev space of order $1$ on $\GG$). 
\noindent  

To introduce the nonlinearity, we define a vector field 
$G=(G_1,\cdots, G_N):\CO^n\rightarrow \CO^n$  acting ``componentwise" as $$G_i(\zeta)=g(|\zeta_i|)\zeta_i\ \  \text{with}\ \ g:\erre^+\rightarrow \erre\ \text{and}\  \zeta=(\zeta_i)\in
\CO^n \ .$$\par\noindent
The vector field $G$ enjoys the important property of gauge (${\rm U}(1)$) invariance, i.e. $G(e^{i\theta} \zeta)=e^{i\theta}G(\zeta)\ .$ 
After this preparation the more common evolution equations on the graph can be defined in the obvious way. 
Main examples are\par\noindent
i) the nonlinear Schr\"odinger equation
\beq
\label{NLSeq}
 i \frac{d}{dt}\Psi_t \ = \ H_{\alpha} \Psi_t +  G(\Psi_t)\  \quad \quad ({\rm NLS\ equation})\ ;   
\eeq
\par\noindent
ii) the nonlinear Klein-Gordon quation
\beq
\label{NLKGeq}
-\frac{d^2}{dt^2}\Psi_t \ = \ H_{\alpha} \Psi_t +  m^2 \Psi_t + G(\Psi_t)\  \quad \quad ({\rm NLKG\ equation})\ .   
\eeq

Notice that from a mathematical point of view the nature of a PDE on a graph amounts to a system of PDE's on suitable finite or infinite intervals (in the case of a star graph $N$ halflines) in which the coupling is given exclusively through the boundary conditions at the vertices.
In the following paragraph we will see several illustrations of this simple remark in the case of our main example, the NLS equation. \par\noindent
A further specification of the model is given specializing the study to the important case of a power nonlinearity 
$$g(z)=\pm |z|^{2\mu},\ \mu >0 \ .$$ 
In the case of NLS equation, the minus sign corresponds to so called focusing nonlinearity and the plus sign to the defocusing one, both meaningful in the applications. In the NLKG equation, the minus sign is the most relevant.
The cubic case is especially important both for the NLS, where it describes the most common phenomenological situations in nonlinear optics and BEC, and in the case of NLKG equation, where it corresponds to the quartic interaction in field theory. 
Simple combination of monomial nonlinearities, such as the physically relevant "cubic-quintic" nonlinearity  
$$g(z)= -|z|^{2} +\epsilon|z|^{4} $$
encountered in nonlinear optics and other fields (see for example \cite{[KA]})
can be considered as well.\par\noindent
In the following we almost exclusively refer to the example of NLS with focusing power nonlinearity and an attractive $\delta$ vertex.
After fixing the model the first preliminar and essential information regards its well posedness, i.e. existence (local or global) and uniqueness of solution in suitable functional spaces. For our preferential model, the NLS equation, the classical line of attack to well posedness is through the integral form of the equation, given by
\begin{equation}\label{intform1}
\Psi_t \ = \ e^{-iH_{\alpha} t} \Psi_0 + i \int_0^t e^{-i
  H_{\alpha} (t-s)}|\Psi_s|^{2\mu} \Psi_s \, ds
\end{equation}
A formal analysis show that for NLS on graphs there exist conserved quantities, similarly to the case of the line or of open set in $\RE^n$.
These are the mass \par\noindent 
\beq\label{mass}
 M[\Psi] = \left\|\Psi\right\|^2 \ ;
\eeq
and the energy 

\beq\label{energy}
 E[\Psi] = \frac12 \|\Psi'\|^2 -\frac1{2\mu+2} \|\Psi\|_{2\mu+2}^{2\mu+2} + \frac{\al}{2}\big| \psi(0)\big|^2 \ ;
\eeq
this last quantity for $\mu\in (0,2)$ is well defined on the domain $\DD(E)=\EE$ of the quadratic part of the energy, i.e. the energy of the linear Schr\"odinger dynamics on the graph.\par\noindent



A sample rigorous result giving well posedness of NLS equation on star graphs and conservation of mass and energy is the following.
\begin{theorem}[Local and global well-posedness in $\EE$]
For any $\Psi_0 \in \EE $, there exists $T > 0$ such that the
equation \eqref{intform1} has a unique solution $\Psi \in C^0 ([0,T),
\EE )
\cap C^1 ([0,T), \EE^\star)$.

\n
Moreover, eq. \eqref{intform1} has a maximal solution $\Psi^{\rm{max}}$
defined on an interval of the form $[0, T^\star)$, and the following ``blow-up
alternative''
holds: either $T^\star = \infty$ or
$$
\lim_{t \to T^\star} \| \Psi_t^{\rm{max}} \|_{\EE}
\ = \ + \infty,
$$
where we denoted by $\Psi_t^{\rm{max}}$ the function $\Psi^{\rm{max}}$ evaluated at time $t$.\par\noindent
Moreover, in the same hypotheses, the following conservation laws hold at
any time $t$:
\begin{equation*}
M[\Psi_t ] \ = M[ \Psi_0 ], \qquad
E[ \Psi_t ] \ = \ E[ \Psi_0 ]\ .
\end{equation*}
Finally, for $0<\mu<2$ and any $\Psi_0 \in \EE $,  the
equation \eqref{intform1} has a unique solution $\Psi \in C^0 ([0,\infty),
\EE )\cap C^1 ([0,\infty), \EE^\star)$.

\end{theorem}
In the previous theorem, $\EE^\star$ is the dual space of $\EE\ .$ 
As in the case of NLS equation on the line, we will call the range of nonlinearities $0<\mu<2$, where existence for all times is guaranteed, {\it subcritical} case, in contrast to the supercritical nonlinearities ($\mu>2$) where blow-up occurs (in the form described in the theorem), or the threshold critical ($\mu=2$) nonlinearity where global existence depends on the size of the initial datum. For details on the proof and generalization to more general couplings at the vertex of the star graph see \cite{[ACFN1], [ACFN3]}. The same theorem holds true for more general nonlinearities with essentially the same proof.
\par\noindent
The previous examples are modeled starting from generators which are nonlinear perturbations of "quantum graphs". \par\noindent
On the other hand, in principle, other equations could be considered. Without embarking here in a general theory we limit to mention some special models. The first one is given by the Benjamin-Bona-Mahony equation (BBM) describing the unidirectional shallow water flow under the long wave and small amplitude approximation, 
$$
 \frac{\partial u}{\partial t} + \frac{\partial u}{\partial x} + u \frac{\partial u}{\partial x} - \frac{{\partial}^3 u}{\partial {x}^2 \partial t} = 0\ .
$$
Of course here the unknown $u$ is real. It is immediate to formally extend such an equation on a graph. To simplify the exposition, consider the case of a star graph with three edges, an Y-junction.\par\noindent 
Let us define a vector $u=(u_1,u_2,u_3)\ $ where $u_i:(0,\infty)\rightarrow \RE$ and suppose that at the vertex the three components $u_i$ satisfy Kirchhoff boundary conditions. These conditions are rather natural in the context of water (and other fluids) waves, corresponding to continuity of the flow and flux balance. \par\noindent
Now orienting the edges of the Y-junction in the outgoing direction as above and setting $\sigma_1=-1\ ,\ \sigma_2=\sigma_3=1 $, let us consider the evolution problem
$$
\frac{\partial u_i}{\partial t} + \frac{\partial u_i}{\partial x} + u_i \frac{\partial u_i}{\partial x} - \frac{{\partial}^3 u_i}{\partial {x}^2 \partial t} = 0\ \quad\quad \ x \in \RE^+\ ,\ \  t > 0;
$$
\begin{equation*}
u_i \in H^2 (\RE^+ )\ ,\ u_1
 (0) = u_2 (0) = u_3 (0)\ ,\ \ \frac{\partial u_1}{\partial x} (0) + \frac{\partial u_1}{\partial x} (0) + \frac{\partial u_1}{\partial x} (0) = 0 \ .
\end{equation*}
This is a system of scalar BBM equations on the halfline, coupled through
the Kirchhoff boundary condition at the origin. In \cite{[BC]} a generalization of this systems to trees (including star graphs) is studied as a model of cardiovascular system, and in particular it is shown that BBM on a tree is well posed. It is interesting that (see \cite{[MuR]}) travelling waves for BBM equation on a tree have been recently constructed for particular vertex conditions. By the way a result of well posedness of NLS on trees in not yet proven, although well plausible; a step in this direction with proof of relevant dispersive estimates is given in \cite{[BI11],[BI12]}. Other nonlinear models of quite different nature are the reaction-diffusion equations on networks, about which some literature exists, in particular regarding pulse propagation in axons and neural networks according to the FitzHugh-Nagumo model and its variants (see \cite{[CMu]} and references therein). Concerning the mathematical setting, a last word should be said about the one dimensional approximation given by the NLS equation on a graph. The graph should be a limit in some sense of a more realistic systems of thin tubes (or guides) connecting at junctions. A first problem is getting the limit of a certain dynamical model, e.g. NLS equation defined on the system of thin guides when transversal size of tube vanishes. It is reasonable to conjecture that this limit should be a NLS on a graph as defined before, but with which boundary conditions? And how the boundary conditions at the vertex could depend on the limit process of shrinking the tubes, guides and junction size? A further difficulty could be the dependence of limit process on  the conditions at the boundary of the thin tube, e.g. Neumann or Dirichlet. These problems have been tackled, with partial solution, in the linear case (see \cite{[DC],[EP],[KZ]} and reference therein). They remain open for nonlinear models, where there is no literature, with the only exception (to the knowledge of the present author) of a series of papers about the reduction of the Ginzburg-Landau equation and its stationary counterpart from thin tubes to graphs (see \cite{[R],[RS],[RSW]} and references therein), where some special boundary conditions at the junction appears in the limit. A second point of view is less phenomenological, and related to the deduction of evolution equation for Bose Einstein condensates from first principle. The dimensional reduction of BEC using scaling trapping potentials is a well understood process in several limiting regimes; in particular a Gross-Pitaevskii energy functional on the line describes the so called cigar-like BECs under certain conditions (see \cite{[LSY], [ANS]} and reference therein). Being confident that a Gross-Pitaevskii equation realizes the correct quasi one dimensional limit, a similar procedure could be attempted on graphlike structure, for example on a Y-junction; to simplify the analysis, one could start directly from a $N$-body theory on the graph, and to attempt at taking the $N\to \infty$ limit with suitable scalings. Notice that the main problem, i.e. the treatment of the boundary condition, it is open also in the simplest case of a graph with two edges, i.e. a line with a defect.
\vspace{5pt}
\par\noindent

\section{The NLS equation on star graphs: rigorous results} 




\par\noindent
After setting the mathematical model and giving the main physical premises and possible applications, we turn to a description of some of its dynamical features.
Again we refer to the case of the focusing NLS equation where more information it is known. Two main topics have been rigorously indagated in the last years: the existence and characterization of standing waves, and the scattering of solitons through a juctions. To introduce the two subjects, let us preliminarly recall some properties of NLS solitons on the line. Let us consider the usual focusing NLS equation with power nonlinearity in one space dimension 
\begin{equation}\label{NLSline}
i \frac{\partial u}{\partial t} (x,t) = -\ \frac{{\partial}^2 u}{\partial x^2}(x,t) - | u(x,t)|^{2\mu}  u(x,t)\qquad
x\in\RE\,,\;t>0\ ;
\end{equation}
as it is well known, it admits a special solution $ u(x,t)=e^{it}\phi$ with
\beq\label{phi}
 \phi (x) =  \lf[ (\mu+1)\ri]^{\f{1}{2\mu}}
{\sech}^{\f{1}{\mu}} (\mu x)\ .
\eeq

A richer family of solution is obtained through application of Galileian and scaling symmetries of the NLS equation:
\beq\label{phigal}
u_{x_0,v,\omega}(x,t) : = e^{i \f v 2 x} {e^{- i t \f {v^2} 4}}e^{i\omega t} {\omega}^{\f{1}{2\mu}}\phi (\sqrt{\omega}(x-x_0-vt))\ .
\eeq
Notice that frequency $\omega$ of the oscillation and amplitude of the solitary wave $u_{x_0,v,\omega}$ are nonlinearly related, and in particular the greater is the amplitude, the greater is the oscillation frequency.\par\noindent
\subsection{Standing waves of NLS on star graphs}
Let us begin with the first subject, standing waves. On the line they appear putting $v=0$ in the previous family of solitary solutions, and they have the character of localized solutions (or "pinned" solitons, in the physical literature) around a certain centre $x_0$. In particular they are the only soliton solutions when travelling waves are excluded by the presence of inhomogeneities; these can be represented for example by external potentials, magnetic fields or, as in our case, by a boundary condition at the junction. In these cases the localization of the standing wave is around stationary points of external potentials, or at the location of singular interactions in case of junctions, point defects, etc. 
Here we will define
standing waves as finite energy solution to a NLS equation (for other models with $U(1)$ symmetry, such as NLKG equation, the definition is exactly the same) of the form \beq
\label{stat-sol}
\Psi_t(x)=e^{i\omega t}\ \Phi_{\omega}(x)\ .  \eeq The function
$\Phi_{\omega}$ is the amplitude of the standing wave. In particular we are interested in standing waves of the NLS equation on graph, equation \eqref{NLSeq}. A regularity argument shows that the standing waves belong in fact to the operator domain of $H_{\alpha}$.
Correspondingly there should exist a frequency $\omega$ and an amplitude $\Phi_{\omega}$ which satisfy in the strong sense the {\it stationary equation }
\beq
\label{stat-eq}
H_{\alpha}\Phi_{\omega} - |\Phi_{\omega}|^{2\mu} \Phi_{\omega} = -\ome \Phi_{\omega}\ .
\eeq
The analysis of this equation on a star graph is simple (see \cite{[ACFN3],[ACFN4]} for details).\par\noindent
On every edge the operator $H_{\alpha}$ coincides with the second derivative, and so it holds
$$\ \ 
-\phi'' - |\phi|^{2\mu} \phi = -\ome \phi \ ;   
$$ 
the most general solution with $\phi\in L^2(\RE^+)$ is (introducing explicitely the dependence on parameters)
\begin{equation}\label{solitonhalfline}
\phi (\sigma, a; x) = \sigma \lf[ (\mu+1)\ome\ri]^{\frac{1}{2\mu}}
\sech^{\frac{1}{\mu}} (\mu \sqrt{\ome} (x-a))\ , \qquad |\sigma|=1 \quad a\in\RE
\end{equation}
so that the components of the amplitudes are
\[
\lf(\Phi_{\omega}\ri)_i=\phi(\sigma_i,a_i)\ ,
\]
where $\sigma_i,a_i$ have to be chosen to satisfy $\Phi_{\omega}\in {\mathcal D}(H_\alpha)\ .$ Notice that on every edge $i$ the stationary state has an amplitude which is "bump-like" or "tail-like" in shape, according to the position of the center $a_i\ $, within the edge $i$ or not.
\par\noindent
Continuity at the vertex implies the following conditions on the parameters of the amplitude  
$$
\sigma_j=1\ , \qquad a_j=\ve_j a\ , \qquad   \ve_j=\pm 1\ \quad j=1,\cdots N\ ,
\  \qquad
a\geq 0 \ .
$$
Being $\sigma_j=1\ \forall j$ we will drop its mention in the following.\par\noindent
Moreover, imposing the $\delta$ vertex boundary condition to $\Phi_{\omega}$ gives
\beq 
\tanh (\mu \sqrt{\ome} a)\sum_{i=1}^N
\ve_i =\f{\al}{\sqrt{\ome}}\ .
\label{eq-a}
\eeq
\par\noindent
so that $\sum_{i=1}^N\ve_i$ must have the same sign of $\al$ if $\al\neq 0$.\par\noindent
Immediate qualitative consequences of the above limitations are:
\begin{itemize}
\item[-] $\alpha>0$ strictly more bumps than tails \par 
\item[-] $\alpha<0$ strictly more tails than bumps\par
\item[-] $\alpha=0$ same number of tails and bumps or $a=0$\par
\item[-] For any value of $\alpha\neq 0 $ there are $[\frac{N+1}{2}]$ states ($[s]$ is the integer part of $s$). \par
\item[-] Lower bound on the allowed frequencies:
$\ \frac{{\alpha}^2}{N^2}<\omega$

\end{itemize}
\begin{figure}
\begin{center}
{\includegraphics[width=.40\columnwidth]{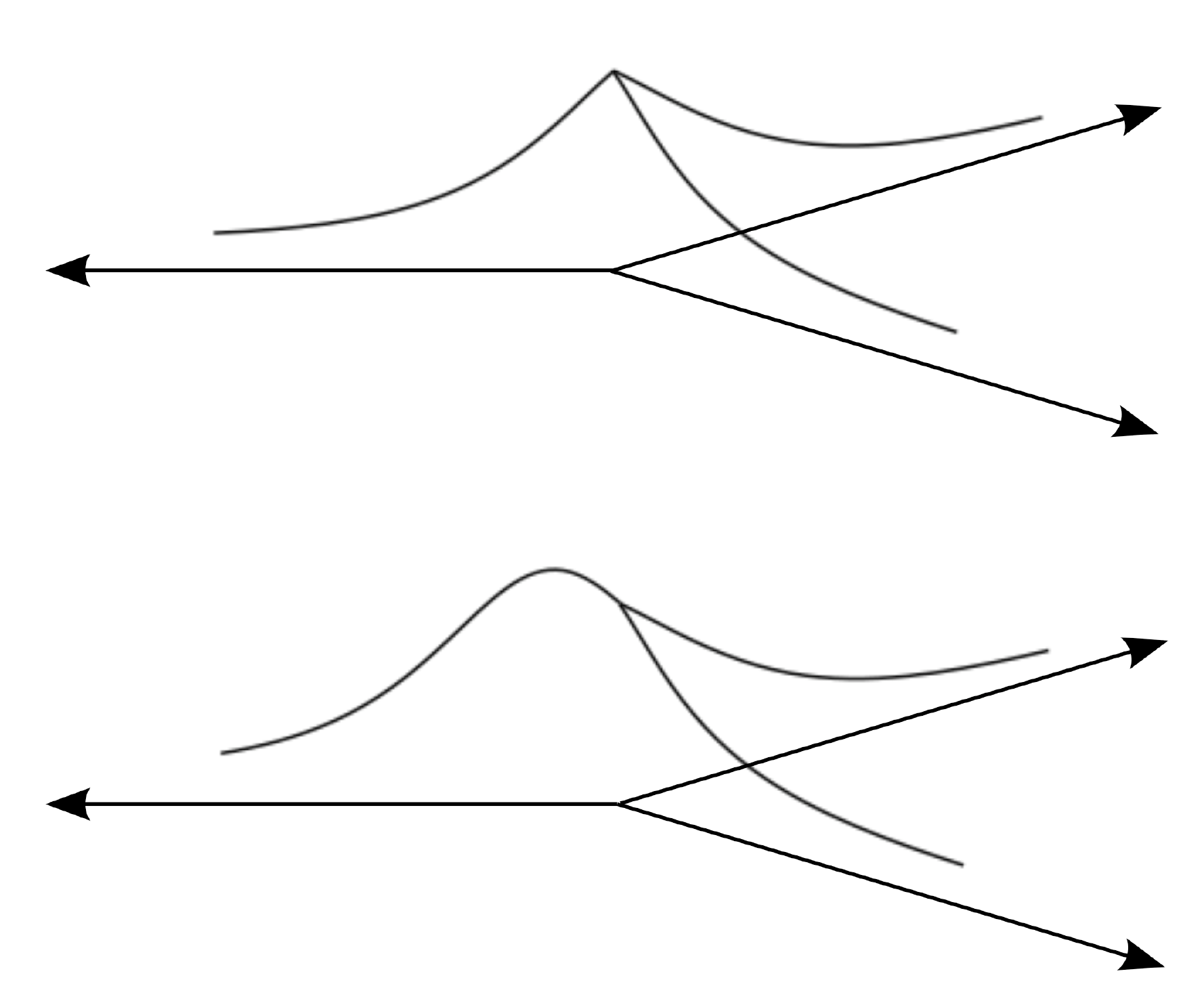}} \quad
{\includegraphics[width=.40\columnwidth]{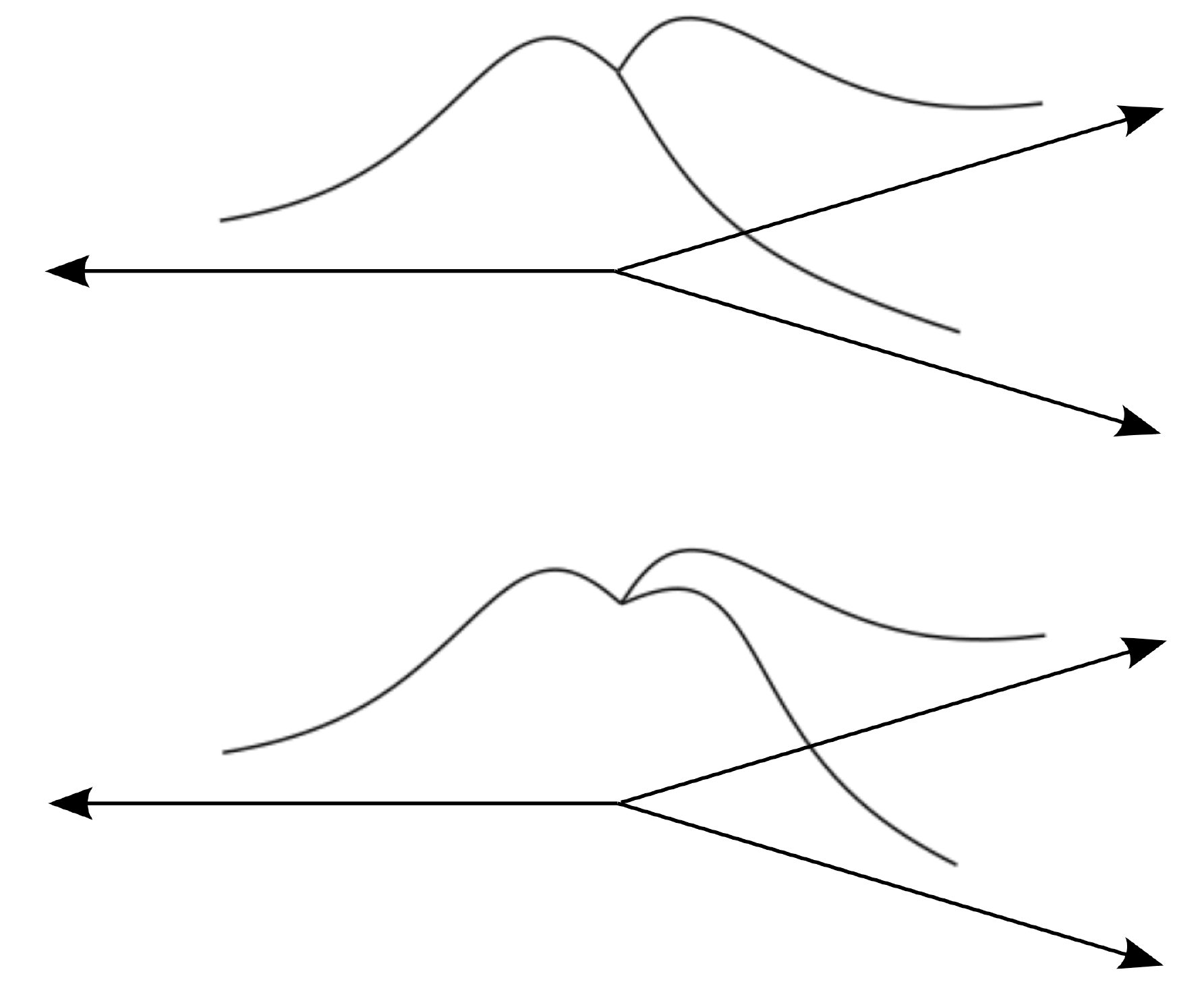}} \\
\caption{$N=3$ stationary states. On the left $\alpha<0$; on the right $\alpha>0$. }
\label{fig:subfig}
\end{center}
\end{figure}
\par\noindent
We index the stationary states $\Phi_{\omega}^j$ with the number $j$ of bumps. With the above limitations this identifies completely the state.
More explicitely the $j$-bumps state $\Phi_{\omega}^j$ is given by
\beq
\lf(\Phi_{\omega}^j\ri)_i(x) = 
\begin{cases}
\phi(a^j;x) & i=1,\ldots j \\
\phi(-a^j;x) & i=j+1, \ldots N
\end{cases} \label{states1}
\eeq 
\beq a^j = \f{1}{\mu \sqrt{\ome}} \,\text{arctanh}
\lf(\f{\al}{(2j-N)\sqrt{\ome}} \ri) \ .\label{states2} 
\eeq 
Concluding, solutions of \eqref{stat-eq} for $\al>0$
are given by $\Phi_{\omega}^j$ with $j=[N/2 + 1], \ldots , N$  and for $\al<0$ by $\Phi_{\omega}^j$ with $j=0 ,\ldots ,[(N-1)/2]$.\par\noindent  
The situation is pictured in Figure 1 for the $N=3$ star graph.\par\noindent
So for $\alpha\neq 0\ ,$ for every $N$ and $\omega>\frac{\alpha^2}{N^2}$ there exist
branches $\{\Phi_{\omega}^j\}$ of stationary states (the branch is unique only in the case $N=2$, i.e. the line). More precisely, the state with $j=0$ arises for $\omega>\frac{\alpha^2}{N^2}$, while to have states with $j>0$ higher frequencies are needed, according to the general relation $\omega>\frac{\alpha^2}{(N-2j)^2}$.

\par\noindent
Now let us consider the Kirchhoff case, $\al=0$. \par\noindent
\par\noindent
From an analysis of the boundary conditions it follows that star graphs with odd or even number of edges behave differently.\par\noindent
For $N$ odd the only value of $a$ compatible with the boundary conditions is $a=0$.
So the stationary state is unique 
   \[
   \left(\Phi_{\omega}^0\right)_i(x) = \phi(0,x)
\quad i=1,\ldots,N
    \]
and it it is composed by N half solitons continuously joined at the vertex.\par\noindent
For $N$ even every real value of $a$ is compatible with the boundary conditions and there is the same number of tails and bumps on the graph.
A one-parameter family of stationary states exists and is given by (notice the slight change of notation used for the Kirchhoff solitary wave only)
\[
\lf(\Phi_{\omega}^a\ri)_i(x) = 
\begin{cases}
\phi(-a,x) & i=1,\ldots N/2 \\
\phi(+a,x) & i=N/2+1, \ldots N
\end{cases} \quad a\in\erre\ .
\]
These stationary states could be thought as $N/2$ identical solitons on $N/2$ lines. The situation is depicted in Figure 2.
As a consequence there exist travelling waves on a Kirchhoff graph with an even number of edges: simply translate the complete solitons on every fictitious line by the same amount $vt$:

$$ 
\Phi_{tr} (t)= e^{i(\frac{v}{2} x-\frac{v^2}{4}
  t+\theta)} \Phi_{\omega}^{a(t)} \qquad a(t)= a+vt\ .
$$

\noindent
\begin{figure}
\begin{center}
{\includegraphics[width=.40\columnwidth]{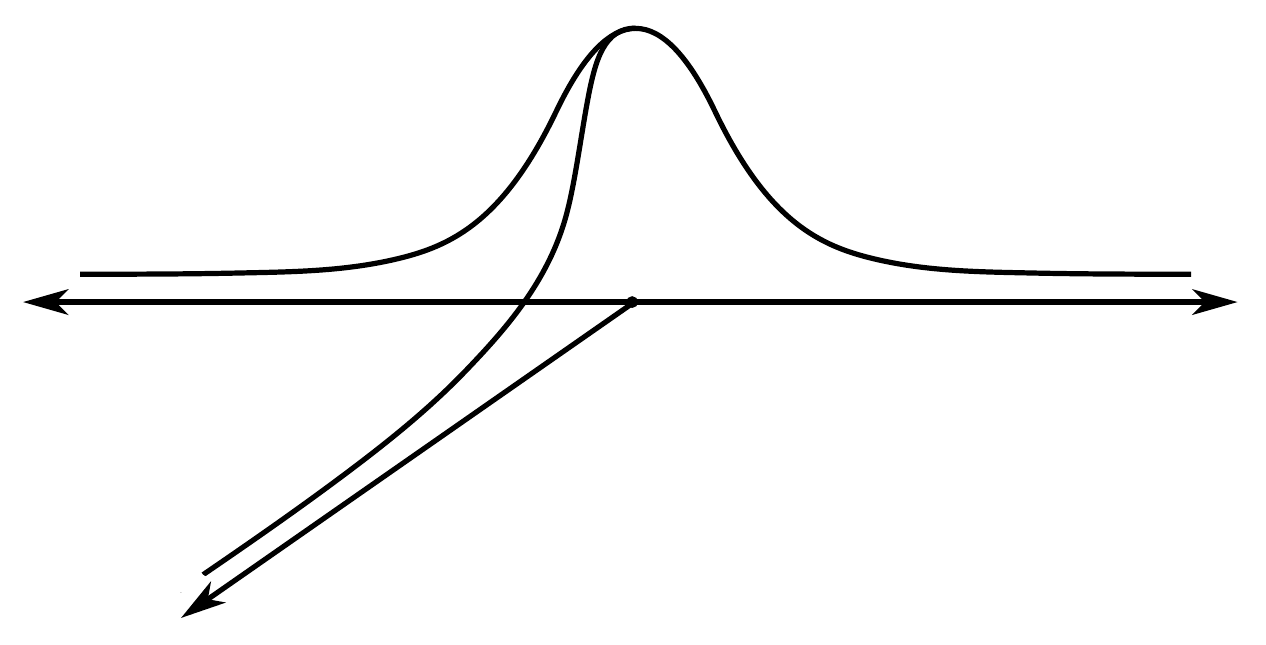}} \quad
{\includegraphics[width=.40\columnwidth]{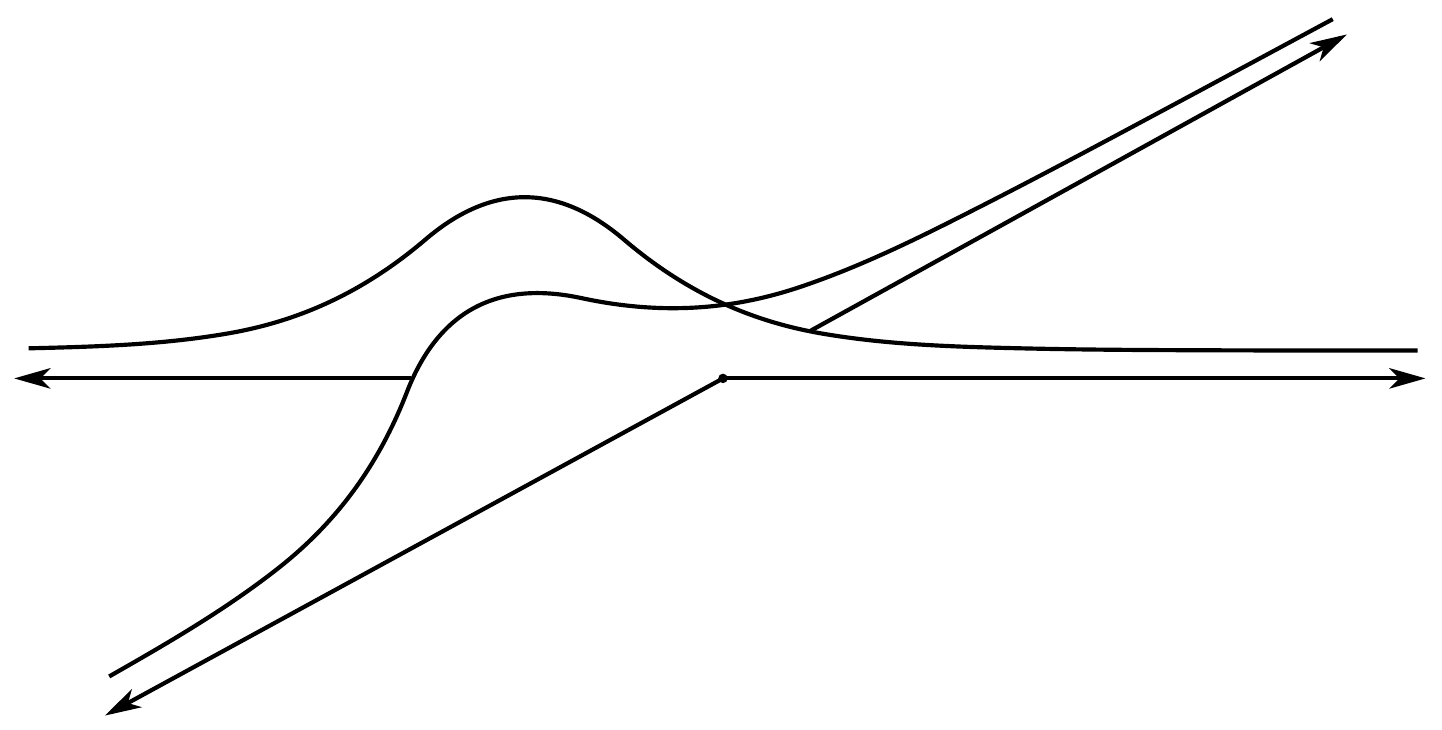}} \\
\caption{The Kirchhoff case $\alpha=0$ for odd (left) and even (right) number of edges. }
\label{fig:subfig}
\end{center}
\end{figure}
\par\noindent
Notice that in the Kirchhoff case stationary states exist for every positive $\omega$.\par\noindent A similar construction can be performed for NLKG standing waves with different restrictions on parameters, i.e. $|\omega|<\sqrt{m^2-\frac{\alpha^2}{N^2}}\ .$ Details will be given elsewhere.
\subsection{Variational properties of standing waves}
After constructing the stationary states, the natural problem is to identify the ground state and, if possible, to order the states in energy, i.e. to describe the nonlinear spectrum of the NLS equation on the graph. This is a variational problem, which is also relevant for the analysis of stability of standing waves. A difficulty immediately arises, in that the NLS energy \eqref{energy} for the focusing NLS equation is unbounded from below, as easily recognized (this is not the case for the defocusing nonlinearity). So the seemingly natural definition of the ground state as the minimizer of the energy is meaningless. Nevertheless the physics of the problems behind the NLS equation suggests that a possible relevant variational problem is a {\it constrained variational problem}: to minimize the energy at fixed mass. With this constraint and for subcritical nonlinearities $\mu<2$ the energy is shown bounded from below for every finite energy state, as it is in the case of NLS equation on $\RE^n$.
In fact the following result holds true (see \cite{[ACFN5]} for details and proofs) for the focusing NLS on a star graph with an attractive delta vertex, $\alpha<0$.
\begin{theorem}[Minimizers for the Energy functional]
\label{t:prob1}
Let $m^\ast$ be defined by
\beq
\label{mstar}
m^\ast=2  \f{(\mu+1)^{1/\mu} }{\mu} \, \lf( \f{|\al|}{N}\ri)^{\f{2-\mu}{\mu}  } \int_0^1 (1-t^2)^{ \f{1}{\mu}-1}\ dt\ .
\eeq
Let $\al <0 $ and assume $m\leq m^\ast$ if $0<\mu < 2$ and  $m < \min\{{m^\ast , \frac{\pi \sqrt{3} N}{4} }\}$ if $\mu=2$ and set
\[
-\nu = \inf \{E[\Phi] \textrm{ s.t. } \Phi\in\mathcal{E}\,,\; M[\Phi] = m\}\ .
\]
Then $0<\nu<\infty$ and there exists $\hat\Phi $ such that $M[\hat\Phi] = m$ and $E[\hat\Phi]=-\nu$.\\
Moreover the minimizer $\hat \Phi$ coincides with the $N$-tail state $\Phi_{\omega_0}^0 $ where $\ome_0$ is such that
$M[\Phi_{\omega_0}^0]=m$.
\end{theorem}
So, for every mass above a certain threshold $m^*$ (which however it is not optimal) the problem of minimizing the NLS energy on the graph at constant mass has a solution if the vertex carries an attractive $\delta$ interaction. More precisely, if the mass constraint coincides with the mass of $N$-tail state, the minimizer is exactly the $N$-tail state.
Some comments are in order.\par\noindent
After some calculation the mass of the stationary states as a function of $\omega$ turns out to be
\begin{align}
 M[\Phi_{\ome}^j] =  \frac{(\mu+1)^{\frac1\mu}}{\mu} \omega^{\frac1\mu-\frac12}  \bigg[(N-2j) \int_{\frac{|\al|}{(N-2j)\omega^{\frac12}}}^1 (1-t^2)^{\frac1\mu -1} dt + 2j  I \bigg]\ ,
\label{e:mome2}
\end{align}
where $I=I(\mu)$ is a certain constant depending on $\mu$ only. We recall that $\Phi_{\ome}^j$ is defined for $\omega\in\left(\frac{|\al|^2}{(N-2j)^2},\infty\right)$. So from \eqref{e:mome2} one easily concludes that 
the functions $M[\Phi_{\ome}^j]$ are increasing in $\omega$ and the minimum value is in correspondence to the threshold $\frac{|\al|^2}{(N-2j)^2}$.
As a consequence the $N$-tail state can have an arbitrarily small mass while the other stationary states have a minimal mass separated from zero. Stated otherways, the manifold $M[\Psi]=m$ for $m<m^\ast$ may not contain all the stationary states, due to the fact that some of them could have too large masses. On the contrary the $N$-tail state always belongs to the constraint manifold since its mass has vanishing lower bound. Taking into account the dependence of $m^\ast$ on $\al$, one concludes that for small $|\al|$ the
constraint manifold contains only the N-tail state while for large $|\al|$ all the stationary states belong to the constraint manifold. Moreover from the expression of $m^\ast$ it follows that to guarantee the existence of ground state for a given mass constraint one has to have a sufficiently deep $\delta$ well. So alternative statements and proofs of the above theorems are obtained fixing $m$ and requiring $\al$ to be sufficiently negative. 
Analogous remarks also apply to the critical case $\mu=2$, the quintic NLS. \par\noindent
With the above precisations a well defined order in energy exists for the stationary states:  
the energies of the stationary states $\Phi_{\omega}^j$, where $\omega=\ome_j$ is such that $M[\Phi_{\ome_j}^j]=m$,  
are increasing in $j$, i.e. they can be ordered in the number of the
bumps (see \cite{[ACFN5]}.\par\noindent At least in one case things are simple: the cubic case.
If $\mu=1$, then $\ome_j$  is independent of $j$ and 
\begin{equation*}
\label{e:omemu=1}
 \ome_j\equiv \ome^* =\frac{ (m + 2|\al|)^2 }{4 N^2} \,; 
 \end{equation*}
\noindent
So, in the cubic case the energy spectrum at fixed mass  can be explicitly computed:
\[
E[\Phi_{\ome^\ast}^j] = -\frac{N}{3} \ome^{\frac32} + \frac13 \frac{|\al|^3}{(2j - N)^2}=-\frac{1}{24} \frac{ (m+2|\al|)^3}{N^2} + \frac13 \frac{|\al|^3}{(2j - N)^2}\,.
\]
\noindent
The energy of the ground state is given by
\[
E[\Phi_{\ome^\ast}^0] = -\frac{1}{24 N^2} m (m^2 + 6m|\al| + 12 |\al|^2 )\ .
\]
\par\noindent
As a final remark, notice that the ground state of the system is the only
stationary state which is symmetrical with respect to permutation of
edges. 
\par\noindent
A second variational problem which has both a physical and mathematical relevance is related to the minimization of the {\it action} functional. 
The action for a nonlinear Schr\"odinger equation is obtained adding the non linear term to the usual action of the linear Schr\"odinger equation
$$
\mathcal{A}[\Psi]=\int_{t_1}^{t_2} \left(\frac{1}{2}\int_{\GG} \Imm(\dot\Psi\ \overline\Psi)\ dx + E(\Psi)\right)\ dt\ .
$$ 
This expression, for a standing wave $\Psi_t=e^{i\omega t}\Phi_{\omega}$ takes the form
\begin{equation*}
\mathcal{A}[e^{i\omega t}\Phi_{\omega}]=(t_1-t_0)\left(E[\Phi_{\omega}]+\frac{\omega}{2} ||\Phi_{\omega}||^2_2\right)\ .
\end{equation*}
This fact suggests to consider the "reduced" action
\begin{equation}\label{action}
S_{\omega}[\Phi]=E[\Phi]+\frac{\omega}{2} M[\Phi] \\
=\frac12 \|\Phi'\|^2 +\frac\ome2 \|\Phi\|^2 -\frac{1}{2\mu+2}\|\Phi\|_{2\mu+2}^{2\mu+2} + \frac{\al}{2}| \Phi(0)|^2 \ ,
\end{equation}
which we continue to call action with a common abuse.\par\noindent
Apart from this lagrangian origin the action $S_{\omega}$ has, at least in the contest of BEC, the physical interpretation of the the grand-potential functional of the condensate corresponding to the chemical potential $\omega$. 
\par\noindent Whatever the theoretical interpretation, the action functional just introduced enjoys the important property that its stationarity condition $S_{\omega}'[\Phi]=0$ coincides with the stationary equation \eqref{stat-eq}.
As for the energy it is easy to see that the action is unbounded from below. Nevertheless a solution of equation \eqref{stat-eq} satisfies necessarily (just take the scalar product of the stationarity equation with $\Phi$) to the constraint
\[
I_\ome[\Phi] = \|\Phi'\|^2 -  \|\Phi\|_{2\mu+2}^{2\mu+2} + \ome \|\Phi\|^2 + \al| \phi(0)|^2 \equiv S'_{\omega}[\Phi]\Phi=0\ .
\]
So it is expedient to search for minima of the action restricted to the above {\it natural} constraint, also called in mathematical literature Nehari manifold. It contains all the stationary states by very definition. Now an immediate calculation show that restricted on the natural constraint the action is
$$ S_{\omega}[\Phi]=\frac{\mu}{2\mu+2} \| \Phi \|_{2 \mu+ 2}^{2\mu +2}\ ,$$ and so it is bounded from below and nonnegative (this is true for every power nonlinearity $\mu$, notice the difference with the constrained energy). The absolute minimum of the action constrained to the Nehari manifold, if existing, is called ground state of the action, as for the energy. Notice that by Lagrange multiplier theorem, the stationary points of the energy at fixed mass are stationary points of the action with Lagrange multiplier ${\omega}$. In fact the ground states of the two problems coincide, as a consequence of the following result (see \cite{[ACFN3]} for a complete discussion and proof).

\begin{theorem}[Minimizers for the Action functional] \mbox{ } \label{mainvar} 
\n
Let $\mu>0$. There exists $\alpha^\ast<0$ such that for $-N\sqrt{\ome} < \al< \al^\ast$
the action functional $S_\ome$  constrained to the Nehari manifold admits an absolute minimum, i.e. a $\hat\Phi\neq 0$ such that $I_\ome [\hat\Phi]=0$
and $S_\ome [\hat\Phi] = {\rm inf} \left\{S_\omega[\Phi]\ :\ I_\omega[\Phi]=0 \right\}$. Moreover, for $-N\sqrt{\ome} < \al <\al^\ast $ the ground state is  $\hat\Phi=\Phi_{\ome}^0$.  

\end{theorem}
\par\noindent
The threshold $\al^\ast$ in the above theorem is known as a function of $N$, $\mu$ and $\alpha$.
The proof of Theorem \ref{mainvar} is quite different from that of Theorem \ref{t:prob1}. Nevertheless the origin of thresholds $\al^\ast$ and $m^\ast$ is analogous, and it relies on the fact that the action of NLS on a Kirchhoff junction has no ground state: the infimum exists, but is not attained at any finite energy state. To explain, let us consider a Y junction, i.e. a $N=3$ star graph. There exist sequences of "runaway" states (see Figure 3) on the Kirchhoff graph given by a complete soliton on a couple of halflines plus a correctly joined soliton tail on the third halfiline; Kirchhoff boundary condition are easily verified, so we have a  sequence of domain elements. Now, the more the big soliton shifts to infinity and the tail estinguishes, the less is the action, which can be shown to converge toward its infimum, strictly lower than the action of any stationary state. The same phenomenon occurs at the constrained energy level as shown in \cite{[ACFN2]}. The bad behaviour of the Kirchhoff junction ($\alpha=0$) prevents the action with $\alpha$ small, or the energy with a big $m$, to have a constrained absolute minimum.
It is a conjecture that the action has a {\it local} constrained
minimum that is larger than the infimum when the condition $-N\sqrt{\ome} < \al< \al^\ast$ fails, and similarly for the constrained energy with a corresponding condition on mass. \par\noindent
\begin{figure}[t!]
\includegraphics[width=8.0cm]{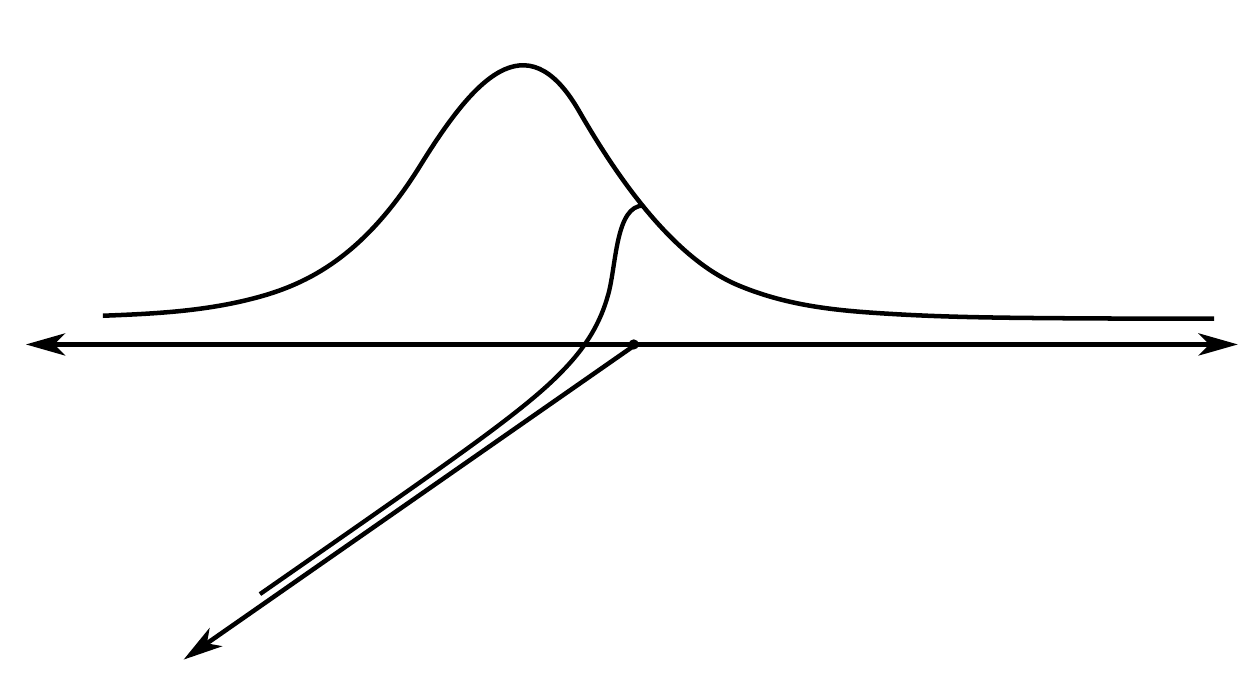}
\caption{A runaway state on the graph}
\end{figure}
\hspace{0.5cm}
\subsection{Orbital stability of standing waves}
Stability is an important requisite of a standing wave, because unstable states are rapidly
dominated by dispersion, drift or blow-up, depending on dynamics, and so are undetectable (stability and instability of NLS with a $\delta$ potential on the line is studied, partly numerically, in \cite{[LeCFFKS]}; see also reference therein). Due to gauge or $U(1)$ invariance of the action and dynamics, a standing wave is not stable in the usual Lyapounov sense. This is a general fact in the presence of symmetries, and it is well known in the example of relative equilibria for finite dimensional mechanical systems. In our case to introduce the main concept af orbital stability, let us consider the special solution $e^{it}\phi(x)$ to equation \ref{NLSline}, where $\phi$ is the initial datum given in \ref{phi}; by scaling and $U(1)$ invariance, $ u(x,t)=e^{i\omega t}{\omega}^{\frac{1}{\mu}}\phi(\sqrt{\omega} x)$ is the solution corresponding to the initial datum ${\omega}^{\frac{1}{\mu}}\phi(\sqrt{\omega} x)$. Choosing $\omega$ close to $1$ the two initial data are close. But their time evolutions are not, because of different frequencies which make distance of solutions to vary with time:
$$
\sup_{t\in \erre}||u(x,t)-e^{it}\phi(x)||_{H^1}>||\phi||_{H^1}\ .
$$
\par\noindent
The same phenomenon occurs on $\erre$ for Galileian invariance: slightly different velocities make travelling solitary waves separate each other. In the case of graphs Galileian invariance is not relevant because it is broken by the junction, travelling waves do not exist and so we concentrate on standing waves.
The point of the previous discussion is that the stability has to be defined {\it up to symmetries}: solutions remain close to the orbit
$e^{i\theta}\Phi_{\omega}$ of a ground state for all times if they
start close enough to it (see for a general discussion \cite{[Caz03], [CL]}). 
The orbit of $\Phi_{\omega}$ is
\[
 {\mathscr O}(\Phi_{\omega})=\{e^{i\theta}\Phi_{\omega}(x),\quad \theta\in \RE\}\ .
\]
\begin{definition}
The state $\Phi_{\omega}$ is orbitally stable if for every $\epsilon>0$ there exists $\delta>0$ such that
\begin{equation*}
\left\{
\begin{aligned}
&\Psi(0)\in\EE\\
&d(\Psi(0), {\mathscr O}(\Phi_{\omega}))<\delta \quad \end{aligned}
\right.
\Rightarrow \quad d(\Psi(t), {\mathscr O}(\Phi_{\omega}))<\epsilon \quad\quad \forall t > 0
\end{equation*}
where 
\begin{equation*}
d(\Psi, {\mathscr O}(\Phi_{\omega}))=\inf_{u\in{\mathscr O}(\Phi_{\omega})}\|\Psi-u\|_{\EE}\ .
\end{equation*}
\end{definition}
\par\noindent
The stationary state $\Phi_{\omega}$ is orbitally unstable if it is not stable.\par\noindent
A general theory of orbital stability was established in the eighties in the classical papers by M.Weinstein \cite{[W1]} and
Grillakis-Shatah-Strauss \cite{[GSS],[GSS2]}, and developed in a number of subsequent papers by many authors; it applies to infinite dimensional Hamiltonian systems (such as abstract NLS equation) when a regular branch of standing waves $\omega\mapsto\Psi_{\omega}$, not necessarily ground states, exists (see for example \cite{[Stuart08],[Ohta]} for recent surveys and results about stability and instability in NLS case). The first step is to give the NLS equation on graph an Hamiltonian structure.
This is achieved in a standard way,  considering an element of $L^2(\GG, \CO)$ as the couple of its real and imaginary part, $\Psi =U+iV \equiv (U,V)$ and endowing the Hilbert space $L^2(\GG, \CO)\equiv L^2(\GG, \erre)\oplus L^2(\GG, \erre)$ so obtained with the real scalar product $\langle \Psi_1, \Psi_2\rangle=\Rea\int_{\GG}\ \overline\Psi_1\Psi_2 = \int_{\GG}\ U_1U_2 + V_1V_2$, and analogously decomposing the higher Sobolev spaces.
The NLS on a graph turns out to be a Hamiltonian system  
\begin{equation}\label{hamiltonian}
\frac{d}{dt} 
\begin{pmatrix}
U \\  V
\end{pmatrix}={\mathcal J} E'[U,V]\ ,\qquad {\mathcal J}=
\begin{pmatrix}
\ 0\ \ I \\ -I\ 0
\end{pmatrix} ,
\end{equation}
where  $E[U,V]\equiv E[\Psi]$ and the functional derivative is defined as usual
\[
 E'[U,V](X,Y)=\frac{d}{d\epsilon}E[(U,V)+\epsilon(X,Y)]_{\epsilon=0}  \ \ \ \forall (X,Y)\in  H^1(\GG, \erre)\oplus H^1(\GG, \erre)\ .
\]
Now linearize the Hamiltonian system around the ground state setting
\begin{equation}\label{perturbation}
\Psi(t)=\left(\Phi^0_{\omega} + W +iZ\right)e^{i\omega t}\ .
\end{equation}
Notice that the previous  definition amounts to pass to a rotating coordinate system, comoving with the ground state. Then the fluctuations $W$ and $Z$ satisfy
 \begin{equation}
\label{linhamsyst}
\frac{d}{dt} 
\begin{pmatrix}
W \\ Z
\end{pmatrix}={\mathcal J} {\mathcal L}
\begin{pmatrix}
W \\ Z
\end{pmatrix}\ ,
\end{equation}
where
\[
{\mathcal L}
\begin{pmatrix}
W \\ Z
\end{pmatrix}=
\begin{pmatrix}
{\mathcal L}_1 W \\ {\mathcal L}_2 Z
\end{pmatrix} 
\]
and ${\mathcal L}_1$ and  ${\mathcal L}_2$ are matrix s.a. operators: their domain coincide, with a slight abuse of notation, with $\DD(H_{\alpha})$ and the action is given by
\beq\label{Luno}
\left({\mathcal L}_1\right)_{i,k} =\left(-\frac{d^2}{dx^2} +\omega - (2\mu+1)|\Phi^0_{\omega,k}|^{2\mu}\right){\delta}_{i,k}\ , 
\eeq

\beq\label{Ldue}
\left({\mathcal L}_2\right)_{i,k} =\left(-\frac{d^2}{dx^2} + \omega - |\Phi^0_{\omega,k}|^{2\mu}\right){\delta}_{i,k}\ .
\eeq
Due to the definition \ref{perturbation}, it turns out that the linearized operator ${\mathcal L}$ coincides with the second derivative of the action: ${\mathcal L}= S''_{\omega}(\Phi_{\omega}^0)$, after identification of the sesquilinear form with the operator via the scalar product.\par\noindent
The first information about stability is given by the spectrum of the operator ${\mathcal L}$. When the linearized operator ${\mathcal L}$ admits at least one eigenvalue with nonvanishing real part the stationary state $\Phi_{\omega}^0$ is said spectrally unstable, otherwise it is said spectrally stable. In fact, due to the presence of conservation laws, in particular of the mass $||\Psi||^2$, the system can be spectrally unstable without being orbitally unstable. 
Precisely, according to the general theory of Weinstein and Grillakis-Shatah-Strauss, for a solitary solution $\Phi_{\omega}$ being orbitally stable is sufficient that:
\begin{itemize}
 \item[$i)$]{\it spectral} conditions hold: 
\begin{itemize}
\item[$i_1)$] $\ker {\mathcal L}_2= \{\Phi_{\omega}\}$ and
the remaining part of the spectrum is positive
\item[$i_2)$] $\ker {\mathcal L}_1= \{0\}$                                  
\item[$i_3)$] the number of negative eigenvalues (the {\it Morse index}) of ${\mathcal L}_1$  is equal to  $1$ 
\end{itemize}
\item[$ii)$]{\it Vakhitov-Kolokolov} condition $\frac{d}{d\omega}\|\Phi_{\omega}\|^2 > 0$
holds. 
\end{itemize}
\par\noindent
These conditions can be verified and one obtains the following theorem
\begin{theorem}
Let $0<\mu \leq2$, $\alpha<\alpha^*<0$, $\omega>\frac{{\alpha}^2}{N^2}$. Then the ground state
$\Phi^{0}_{\omega}\ $ is orbitally stable in $\\E\ .$ Moreover, if $\mu>2$ there exists $\omega^*$ such that $\Phi_{\omega}^0\ $ is orbitally stable for 
$\omega\in (\frac{{\alpha}^2}{N^2},\omega^*)\ $ and orbitally unstable for $\omega>\omega^*\ .$ 
\end{theorem}
\par\noindent
The theorem gives orbital stability of the ground state for every $\omega$ also for the critical nonlinearity $\mu=2\ .$  The case $\omega=\omega^*$ is undecided. The proof is a calculation for the Vakhitov-Kolokolov condition; and concerning the spectral conditions, the Morse index of ${\mathcal L}_1$ is one as a consequence of the fact that the action has a minimum at $\Phi_{\omega}^0\ $ restricted to the codimension one Nehari manifold. The other details are in \cite{[ACFN3]}. \par\noindent
We end this subsection with some comments. There is a second strategy to show orbital stability, which again makes use of a variational property. A stationary state which minimizes the energy at constant mass is orbitally stable (see the classical paper \cite{[CL]}, where several example are treated). So a direct consequence of Theorem \ref{t:prob1} is orbital stability of ground states. Nevertheless some remarks are in order. The first is that while Theorem \ref{t:prob1} and in general the concentration compactness technique developed in \cite{[CL]} give information only if absolute constrained energy minima (i.e. ground states) exist, the above Weinstein and Grillakis-Shatah-Strauss theory is more general, and it is in principle applicable to every stationary state of the action, for example excited states of NLS equation on star graphs above described.
For excited states the expectation is orbital instability, which is in fact the case, as it will be shown elsewhere.
The difficult part of the analsis is the calculation of the Morse index of operator ${\mathcal L}_1$  and the use of general results in \cite{[GSS2]} and their recent refinements in \cite{[Ohta]}. One could ask if a similar analysis could be performed on more complex graphs, for example trees or graphs with loops. Of course the fact that stationary states on star graphs are completely known, which is a rare case, is a strong facilitation in obtaining precise results. In the case of less trivial graphs it is generally impossible to obtain explicitly standing waves, but some simple non trivial graphs can be probably treated along the lines discussed. On more general grounds, when the linear part of the model, i.e. the underlying quantum graph has an eigenvalue, for example corresponding to the linear ground state, bifurcation theory suggests that a branch of nonlinear standing waves exists and it bifurcates from the vanishing solution in the direction of the linear ground state (see \cite{[RW]} for a classical application to the case of NLS with external potential). But there are several problems which arise at this point. The first is that a direct analysis of the conditions guaranteeing orbital stability or instability of standing waves, in particular counting of negative eigenvalues of ${\mathcal L}_1$ and verification of Vakhitov-Kolokolov condition, becomes in general impossible or at least very difficult. A second problem is that bifurcation theory allows to identify branches of nonlinear stationary states which have a linear counterpart, but how to obtain branches of states without linear counterpart, which exist as we know from the example of nonlinear excited states in star graphs? A guess is that excited or in general bound states without linear counterpart bifurcate from (not small) solitary waves turning on the external potential. A different extension could be in the direction of different boundary conditions at the vertices. In such a case one expects new dynamical effects, for example bifurcation and symmetry breaking of ground states as proved for the NLS on the line with $\delta'$ interaction (see\cite{[AN12]}). A final open and difficult problem is the so called asymptotic stability of standing waves. In our case a standing wave $\Phi_{\omega}e^{i\omega t}$ is said asymptotically stable if for every solution $u(t)$ starting near $\Phi_{\omega}$ in the energy norm, one has the representation
\begin{equation}
u(t) = e^{i\omega_{\infty} t} \Phi_{\omega_{\infty}}
+U_t*\Psi_{\infty} +R_{\infty}(t)\ , 
\end{equation}
where $U_t$ is the unitary evolution of the linear s.a. operator $H_{\alpha}$, and  $\Psi_{\infty}$,
$R_{\infty}(t) \in L^2(\GG)$, with $\| R_{\infty} (t)\| =
O(t^{-\beta})$ as $t \rightarrow +\infty \ ,$ for some $ \beta>0$ and $\omega_{\infty} > \frac{\alpha^2}{N^2}\ .$ So every solution starting near an asymptotically stable standing wave is asymptotically a standing wave (not necessarily the original one) up to a remainder which is a sum of a dispersive term (a solution of the linear Schr\"odinger equation) and a tail small in time. The physical interpretation of the concept is that dispersion, or radiation at infinity, provide the mechanism of stabilization, or relaxation, toward the asymptotic standing wave or more generally solitons. See for example \cite{[GS]} which treats NLS equation with a potential on the line, and references therein. The problem is very hard and there is only partial information also in the case of a NLS equation on the line with a $\delta$ potential (see \cite{[DP]}). Part of the difficulties are due to the fact that the Hamiltonian structure plays a role in the analysis and this makes unavoidable to study the spectrum of the not selfadjoint and not skew adjoint operator ${\mathcal J} {\mathcal L}$ (the Hamiltonian linearization) and to get dispersive estimates about its evolution ${\rm exp}({t {\mathcal J} {\mathcal L}})$; this introduces some interesting and new mathematical problems about operators on graphs, already at the linear level. Moreover the proof of possible asymptotic stability requires a control of the decay of nonlinear remainders; this control depends in a critical way on several analytical tools (in particular dispersive Strichartz estimates) which at present work only for restricted classes of nonlinearities; for example, for subcritical power nonlinearities the procedure fails.


\subsection{Scattering of fast solitons on junctions}
In the preceeding section the behaviour of localized standing soliton solutions of the NLS equation and of the solutions in their vicinity has been studied. Here, concentrating on the case of cubic NLS equation on a three edge star graph, we explorate a different region of state space of this model, that of the asymptotically travelling waves. 
\par As recalled before among the family of solitary solutions of NLS equation on the line given by the action of the Galilei group on the
elementary function
\begin{equation}\label{solzero}
 \phi (x) \ = \ \sqrt 2 \cosh^{-1} x\ ,\qquad x\in\RE\,,
 \end{equation}
 there are the translating waves
\begin{equation}\label{solt}
 \phi_{x_0,v}(x,t) = e^{i \f v 2 x} {e^{- i t \f
    {v^2} 4}}e^{it} \phi ( x - x_0 - vt)\,\qquad x\in\RE\ ,\ t\in\RE\ ,\ v\in\RE.
\end{equation}
\par
This special state, when put on a single edge and pushed to infinity, could be reasonably considered as an asymptotic soliton traveling on the graph. Of course the presence of the vertex breaks galileian invariance and the soliton cannot rigidly translate in the course of evolution, also for the simplest graph, i.e. the star graph with Kirchhoff boundary condition at the vertex. The classical and well known algebraic and analytic techniques to construct exact solutions of (integrable) cubic NLS on the line fail on a graph. The interaction with the Y-junction could be in general quite complex, and from a mathematical point of view essentially nothing is known, if not in the special case where the asymptotic solitary wave is a {\it fast} soliton, in a sense that will be made precise later. In such a case, after the collision of a soliton with the vertex there exists
a time lapse during which the dynamics can be described as the
scattering of three splitted solitary waves, one reflected on the same
edge where the soliton was resident asymptotically in the
past, and two transmitted solitary waves on the other edges. The amplitudes of the reflected and transmitted
solitary waves are given by the scattering matrix of the linear
dynamics on the graph. This behaviour takes place with a small error along a time scale of the order $\ln v$ after the collision, where $v$ is the asymptotic velocity of impinging soliton. On this time scale, Figure $4$ is a realistic approximation of the process. 

\begin{center}
\begin{figure}[h!]
\begin{minipage}{0.47\textwidth}
\center{$t=0$}
\end{minipage}
\begin{minipage}{0.47\textwidth}
\center{$t$ large}
\end{minipage}

\includegraphics[width=0.9\textwidth]{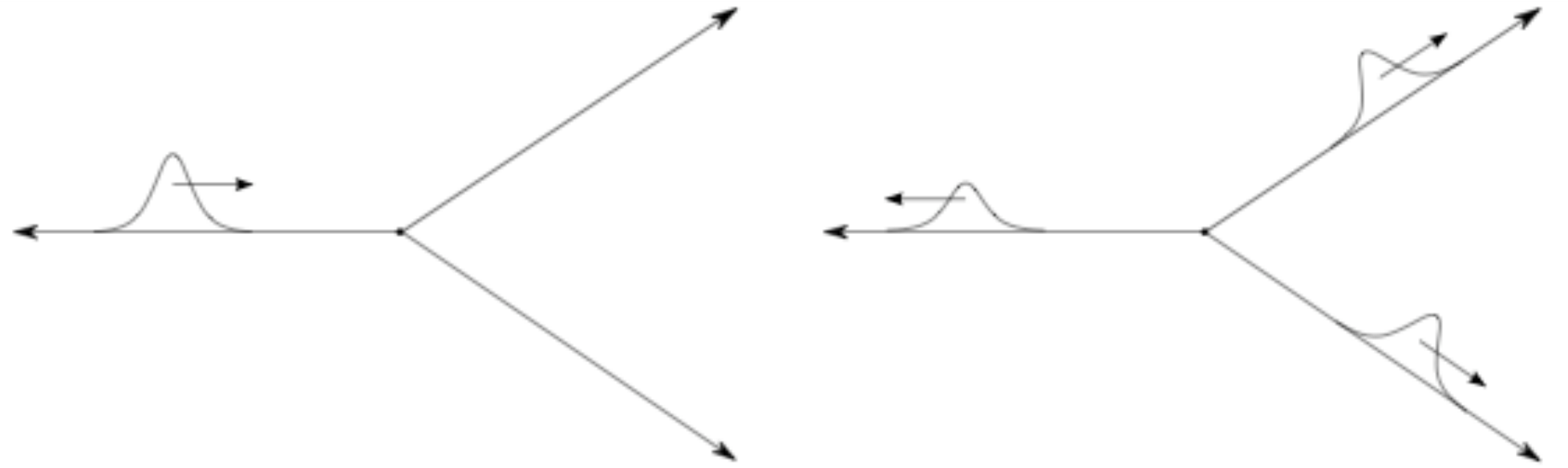}
\caption{Scattering of fast soliton on a Y-junction}
\end{figure}
\end{center}
\vspace{10pt}
The results in \cite{[ACFN1]} and here discussed are inspired by the analogous analysis for NLS equation on the line with a repulsive $\delta$ potential in the paper \cite{[HMZ07]}. It should be said that a graph with two edges is equivalent to a line with a point interaction, so the treatment in \cite{[ACFN1]} where several types of vertices are considered (Kirchhoff and repulsive $\delta$ and $\delta'$, but the analysis could be extended to more general boundary conditions) shows how to generalize the results of \cite{[HMZ07]} to other point interactions on the line and star graphs.
To simplify the exposition, let us consider the following setting
\begin{itemize}
\item Cubic NLS (this is essential: see later)
\item Kirchhoff  vertex $H_{\alpha}=H$ (more general boundary conditions are allowed)
\item Initial state ($v\gg 1$; $x_0 \geqslant v^{1-\delta}$, with $0<\delta < 1$; $\chi$ a smooth cutoff of the tail at the vertex)
\beq\label{insol}
 \Psi_0 (x)  = 
 (\sqrt2\chi (x) e^{-i \frac v 2 x} \cosh^{-1}( x - x_0),0,0)
\eeq
\end{itemize}
\par\noindent
We are interested in the evolution $\Psi_t$ of this
initial condition. To this end, we will find an approximate solution of the equation:
\beq\label{intKirchh}
\Psi_t \ = \ e^{-iH t} \Psi_0 + i \int_0^t e^{-i
  H (t-s)}|\Psi_s|^2 \Psi_s \, ds\ .
\eeq
\par\noindent
The dynamics can be divided into three phases.\par\noindent
The first phase is the approach to the vertex in the time interval $t \in [0, t_1]$, where $t_1 = x_0 / v - v^{- \delta}\ .$
In this phase the incoming (quasi) soliton moves from $x_0$ to $x_0-v t_1=v^{1-\delta}$ and ends the run at a distance of order $v^{1-\delta}$ from the vertex.
During this phase only a small tail of the pulse touches the vertex, the solution $\Psi_t$ behaves much as the solitary solution of the NLS in $\RE$ and it remains supported on the edge $e_1$ with an exponentially small error.
Choosing as the approximating function 
$
\Phi_t (x)\ = (\phi_{x_0,-v}(x,t)
,0,0)$
the following estimate (in $L^2$ norm: control of masses) holds true
\begin{lemma}
For any $t \in [0, t_1]\ ,$   
$\ \ 
\| \Psi_t - \Phi_t \| \leqslant C e^{-v^{1 - \delta}}
$ for $v\gg1$  and  $\delta\in(0,1)$.
\end{lemma}
The proof consists in the accurate use of the well known and already cited Strichartz estimates (an ubiquitous tool in the study of nonlinear dispersive equations: see \cite{[Caz03]}) to control distance between
the unperturbed NLS flow and the NLS flow on the graph.\par\noindent
The second phase is the interaction phase, when the ``body'' of the soliton  crosses the vertex. It occurs during the time $t \in [t_1, t_2]$, where $t_2 = x_0 / v + v^{- \delta}]$.
The time interval $t_2-t_1$ is small (of order $v^{-\delta}$, $0<\de<1$), and the effect of the nonlinear term is demonstrably small.
The soliton is fast and the pulse travels for  a large distance (of order $v(t_2-t_1)=v^{1-\delta}$, $0<\de<1$). This allows the linear dynamics being described by using a scattering approximation. Let $\tr$ and $\rf$ be the transmission and reflection coefficients of the Kirchhoff interaction $H$, 
$\tr  =  \frac  {2} {N}$ and  $\rf  =  -\frac  {N-2} {N}\ .$
This means that the function $
\Psi(k,x_1,x_2,x_3)=(e^{-ikx_1}+\rf e^{ikx_1}, \tr e^{ikx_2}, \tr e^{ikx_3})
$
satisfies the Kirchhoff boundary conditions, and it is a solution of $H\Psi=k^2\Psi$ in distributional sense.
The approximating function is chosen then as (recall \ref{solt})
\[
\Phi^{S}_t=
\big(\phi_{x_0,-v}(t) +
 \rf\,\phi_{-x_0,v}(t),
 \tr\,\phi_{-x_0,v}(t),
 \tr\, \phi_{-x_0,v}(t)\big)
\] 

\begin{figure}[h!]
\begin{center}
\includegraphics[width=0.95\textwidth]{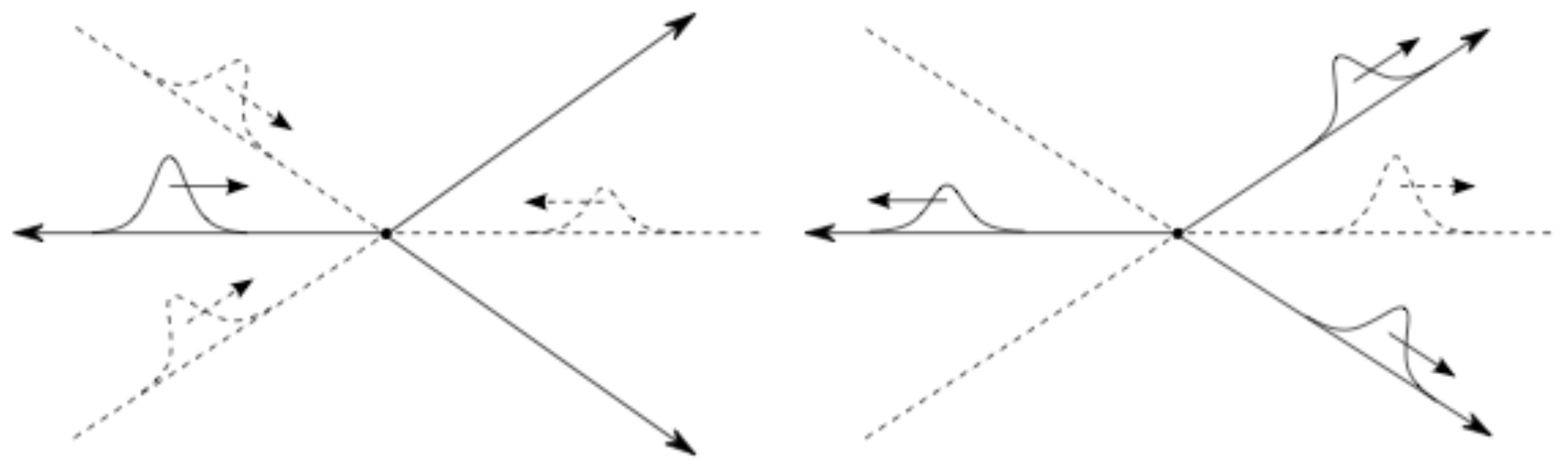}
\end{center}

\caption{On the left side the state $\Phi^{S}$ at time $t_1$. On the right side the state  $\Phi^{S}$ at time $t_2$.
Each edge of the graph is extended to a line by ideally adding the half line $(-\infty, 0]$ represented by the dashed lines.}
\end{figure}	   
The choice of this reference approximate dynamics is explained in Figure 5. Reflected and trasmitted contributions are represented as tails (with $\rf$ and $\tr$ factors) of travelling solitons on "ghost" halflines. This trick is useful and has the further advantage to introduce fictitious lines where some representations and known properties of solitary waves are at disposal (see \cite{[ACFN1]} for details). In  any case, the true solution is compared with this approximate solution with an error small in an inverse power of velocity:
\begin{lemma}
For any $t \in [t_1, t_2]$ $
\| \Psi_t - \Phi^S_t \| \leqslant C v^{-\delta / 2}
$
for $v\gg1$  and  $\delta\in(0,1)$.
\end{lemma}
The proof is technical and makes use of various well chosen representation for the linear time evolution in the interaction phase, an integral equation for the difference between the true solution and the approximate one, and an iteration of Strichartz estimates for evaluating errors.\par\noindent
Finally there is the post interaction phase, in the time interval $t \in [t_2, t_3]$, with $t_3=t_2+T \ln v$ where the free NLS dynamics dominates again; however, now the initial data are not asymptotic solitary waves, but waves
with approximately soliton-like profiles and wrong amplitudes, due to the presence of scattering coefficients. 
Precisely, for $t\geqslant t_2$ and $x\in\RE$ we define the functions $\phi^{tr}$ and $\phi^{ref}$ by

\begin{minipage}{0.48\textwidth}
\[
\left\{
\begin{aligned}
&i \pd{}{t}\phi^{tr} =-\pd{^2}{x^2}\phi^{tr} - |\phi^{tr} |^2\phi^{tr}\\ \\
&\phi^{tr}(x,t_2)=\tr\,\phi_{-x_0,v}(x,t_2)
\end{aligned}
\right.
\]
\end{minipage}
\begin{minipage}{0.48\textwidth}
\[
\left\{
\begin{aligned}
&i \pd{}{t}\phi^{ref} =-\pd{^2}{x^2}\phi^{ref} - |\phi^{ref} |^2\phi^{ref}\\ \\
&\phi^{ref}(x,t_2)=\rf\,\phi_{-x_0,v}(x,t_2)
\end{aligned}
\right.
\]
\end{minipage}

The approximate solution is defined as
\[
\Phi_t^{out}:=
\Big( \phi^{ref}(t), \phi^{tr}(t), \phi^{tr}(t)\Big) 
\]
and the it satisfies the following lemma, which is the main result. 
\begin{lemma}
Fix $T>0$, then for any time $t\in[t_2,t_2+T \ln v]$, there exists $0<\eta<1/2$ such that 
\[
\|\Psi_t-\Phi^{out}_t\|\leqslant Cv^{-\eta}
\]
\end{lemma}
To prove the lemma one has to use in an essential way the fact that the cubic NLS equation is integrable on the line, and thanks to this it is possible to get large time behaviour of initial "lowered solitons" along the nonlinear evolution (see \cite{[HMZ07]} which is the source of the idea). Again with due estimates of the errors these results can be translated on the graph, ending with the quoted result.
\par\noindent
We end with some remarks and comments. \par\noindent
The linear Hamiltonians in \ref{intKirchh} to which the
theorem refers, when different of the Kirchhoff one, have to be rescaled in order to give a nontrivial
scattering matrix in the regime of high velocity; for example in the case of a delta potential one has to set $\alpha \to v\alpha\ .$\par\noindent
The estimates are in $L^2$-norm. So the present analysis of soliton scattering is rigorous for what concerns mass transmission and reflection. On the contrary, the result proposed in \cite{[HMZ07]} is in $L^\infty$-norm and the control is on the profile of the outcoming pulses. \par\noindent
In the time interval $t_2 < t < t_2 + T\ln v\ ,$ the last lemma implies that
$
\frac{\|\Phi^{out,j}_t\|}{\|\Psi_t\|}=|\tr| + {\mathcal{O}}(v^{-\sigma})\ \ (j=2,3)
$
\par\noindent 
holds for a certain $\sigma>0$ and where $\tr$ is the transmission coefficient. So, in the limit of fast solitons, i.e. $v\to \infty$, the ratio which defines the nonlinear scattering coefficient converges to the corresponding linear scattering coefficient. Analogously for reflection coefficient.\par\noindent
The results discussed in this section are given and proved for a repulsive interaction (no negative eigenvalues) at the vertex.
In the presence of eigenvalues a more refined analysis has been performed in \cite{[DH]} for the NLS equation on the line with an
attractive $\delta$ potential, giving results qualitatively similar to the repulsive case. Notice that this is the case in which stable standing waves exist.\par\noindent In view of the previous remarks, the directions in which the scattering of solitons on star graphs could be extended with a certain amount of technical work, but without introducing new ideas are: a)the (straightforward) case of NLS on star graphs with more than three edges; the case in which the underlying linear quantum graph with $\delta$ or $\delta'$ vertex has bound states; with some caution, the case of general selfadjoint boundary conditions. Nothing it is presently known about the possible existence of multisolitons, or about the collision of several solitons at the vertex. \par\noindent
On the contrary, scattering of solitons on graphs having a less trivial topology is an open problem. In particular it could be interesting the study of propagation of solitons on trees, about which something has been said before in relation to BBM equation. Notice that among the main technical ingredients are Strichartz estimates, and every generalization needs their validity. Recent advances in this direction are given in \cite{[BI11],[BI12]}. To give an idea of the interplay between nonlinearity and scattering on complex networks, let us briefly discuss the interesting paper \cite{[GSD]}. The authors consider a (possible complex) graph with bounded edges where NLS dynamics is posed, and to this localized nonlinear network two external legs are attached where linear propagation occurs. 
After some general remarks, numerical results are given and discussed concerning the scattering of stationary wave $a_{in}e^{ikx}$ incoming from one of the external edge, entering the nonlinear network and outgoing from the second external edge: numerical experiments are done for a large spectral range of $k$'s and intensities $I_{in}=|a_{in}|^2\ .$ It turns out that also for relatively simple examples of networks, scattering is dominated by sets of sharp resonances of the underlying linear model; these tend to sensibly amplificate the effect of nonlinearity and they prevent to consider it as a small perturbation. This moreover gives rise to typical effects in nonlinear dynamics, such as multistability and hysteresis. The resonances correspond to longliving states captured in the networks, and they are a result of the non trivial topology of the graph. In this sense the effect should be considered as a general phenomenon which could be expected in real experimental settings involving nonlinear networks.\par\noindent
Another important problem concerns the possibility of extending the timescale
of validity of approximation by the solitary outgoing waves. In a different model (scattering of
two solitons on the line) in \cite{[AbFS]},
some considerations are given about the possibility of longer timescales of outgoing soliton approximation depending on the initial data and
external potential, but it is unclear whether similar
considerations can be applied to the present case.
\par\noindent
As regards the limitation to cubic NLS,  the fundamental asymptotics proved
in \cite{[HMZ07]} and used to get the analogous result on graphs depend on the
integrability of cubic NLS. One can conjecture that
for nonlinearities close to integrable the outgoing waves are close to solitons over timescale similar to the above ones. In this respect it appear as interesting the recent result of Perelman on
the asymptotics of colliding solitons on the line (\cite{[P1]}
). \par 
As a final remark, let us notice that the two main dynamical features here described and about which a description, if partial, has been achieved, i.e. standing waves and their neighborhood on one hand, and scattering of fast solitons on the other hand, correspond to states and regimes far apart in the energy space of the system. Many interesting physical phenomena experimentally well tested are probably in a different or intermediate region. In particular capture and generation of solitons are not yet understood from a rigorous point of view, and they represent a challenge to mathematical methods and theoretical interpretation where perhaps the simple but non trivial model of NLS equation on graphs could offer some insight.

\vskip5pt {\bf Acknowledgements.}
\par\noindent 
The author is grateful to his friends and collaborators Riccardo Adami, Claudio Cacciapuoti and Domenico Finco, with whom a large part of the material here described was developed in the last years. Special thanks are due to Valeria Banica, Gianfausto Dell'Antonio, Reika Fukuizumi, Sven Gnutzmann, Peter Kuchment, Delio Mugnolo and Uzy Smilansky for useful discussions.



\end{document}